# Beyond the Chatbot: Co-Learning and Co-Teaching through a Dual-Persona Generative-AI Assistant

Chaido Mizeli[1], Marina Delianidi[2], Konstantinos Diamantaras[3]
[1]International Hellenic University, Thessaloniki, Greece — cmizeli@ihu.edu.gr
[2] International Hellenic University, Thessaloniki, Greece — dmarina@ihu.gr
[3]International Hellenic University, Thessaloniki, Greece — kdiamant@ihu.gr

**Abstract—**In this paper we present a generative-AI application developed to support both teachers and students in secondary education. The system employs two Large Language Models (LLMs) – Gemini and DeepSeek, and a Small Language model (SLM) – Gemma, integrated within a Retrieval-Augmented Generation (RAG) framework to create a pedagogically grounded, Greek-language assistant capable of adapting its reasoning and communication style to the user's role. Unlike conventional chatbots, the assistant introduces *Pedagogical Persona Switching*, a dual-role mechanism that enables the same AI model to act as both a teaching companion and a learning guide. When interacting with a student, the assistant produces accessible, curriculum-grounded explanations that promote conceptual understanding while remaining faithful to the officially approved educational sources. When addressing a teacher, the assistant generates diverse artifacts, including assessment items, classroom activities, and structured lesson plans. This dual-persona approach bridges knowledge retrieval with pedagogical creativity, enhancing differentiated learning. Utilizing a RAG paradigm tailored to the Greek educational domain, the architecture segments official textbooks into coherent units. Enriched with specific metadata, these units preserve curricular structure and instructional context, demonstrating how generative AI optimizes modern instructional design.

These units are represented by domain-specific sentence embeddings fine-tuned on Greek semantic similarity tasks. This linguistic adaptation ensures precise meaning alignment between user queries and content segments, maintaining fidelity to the official curriculum. All representations are organized within a retrieval index, guaranteeing transparent and source-grounded generation.

The initial case study focuses on Home Economics in Greek lower-secondary education (Grades A–B), a cross-disciplinary subject that integrates elements of economics, health education, and social responsibility. The assistant has been developed to support both learners and educators in complementary ways. In future classroom implementations, students will be able to use it to clarify key concepts such as financial literacy, resource management, and healthy living, while teachers could employ it to design authentic instructional materials, formative assessments, and classroom activities aligned with the official curriculum. Home Economics (HE) was chosen as the pilot domain because it is a non-STEM, value-oriented subject focused on early adolescents, emphasizing ethical reasoning, social awareness, and everyday decision-making. This unique context allows us to explore generative AI's pedagogical potential beyond traditional core subjects, particularly its capacity to foster reflective reasoning and civic responsibility in young learners.

The project introduces an innovative generative framework combining three key elements: (a) a localized, curriculum-specific Retrieval-Augmented Generation (RAG) environment, (b) semantic educational text segmentation, and (c) adaptive persona-driven prompting. More than just a technical solution, this framework demonstrates how generative AI can be pedagogically and linguistically aligned with national curricula and local educational contexts. The resulting model offers a reproducible foundation for future AI integration into authentic learning settings, supporting teacher agency, learner engagement, and values-based secondary education. The study elevates the concept beyond a simple chatbot, proposing a structured, contextually adaptive framework for pedagogical generative assistants that effectively bridge technology, curriculum, and human learning.

## 1. Introduction

The rapid evolution of Generative Artificial Intelligence (GenAI) is fundamentally reshaping the educational landscape, moving past simple digital assistants toward sophisticated systems capable of complex content creation (Kuhail et al., 2023). While first-generation chatbots focused primarily on quick information retrieval (Kuhail et al., 2023), modern Large Language Models (LLMs) enable the development of tools that can understand, generate, and predict human-like language for advanced pedagogical tasks. Recently, the emergence of Small Language Models (SLMs) has offered a more compact and resource-efficient alternative (Touvron et al., 2023). These models feature fewer parameters and require significantly less computational power, making them ideal for specialized educational applications where resources are constrained.

Despite these advancements, a significant challenge remains: bridging the gap between raw knowledge retrieval and the nuanced needs of classroom instruction. Traditional chatbots often lack the contextual grounding required to remain faithful to specific national curricula. To address this, the Retrieval-Augmented Generation (RAG) paradigm has emerged as a crucial solution (Lewis et al., 2020).

RAG enables a model to search a specific, external knowledge base to retrieve factual, up-to-date information before generating a response. This approach significantly reduces hallucinations and ensures that answers are grounded in real, verified data (Lewis et al., 2020).

This article introduces S.T.A.R.T. BOT (Student Tutor Adaptive Role Tool Bot), a generative AI application specifically tailored to the Greek education sector. The system utilizes a RAG framework to ensure that AI responses are strictly based on official school textbooks[1]. Through this architecture, formal learning materials are semantically segmented into coherent units enriched with metadata, ensuring that the curriculum structure and teaching framework are maintained.

The key innovation of S.T.A.R.T. BOT lies in its Pedagogical Persona Switching mechanism. This dual-role approach allows the same AI model to act as both a teaching assistant for educators and a learning guide for students. By utilizing persona-based prompting techniques (Zamfirescu-Pereira et al., 2023), the assistant generates instructional materials for teachers—such as assessment items and lesson plans—while providing curriculum-based explanations and feedback to students. Our research focuses on a pilot study in Home Economics, examining whether SLMs can effectively compete with LLMs in providing safe, reliable, and curriculum-aligned educational support.

Thus, the research questions we pose in this paper are:

1. Can Small Language Models compete with Large Language Models in teaching and learning applications?
2. Can an AI Digital Assistant rely on Small Language Models?

## 2. Related Work

[1] https://ebooks.edu.gr/ebooks/

The landscape of educational technology is currently undergoing a transformative shift, precipitated by the rapid progression from static information retrieval tools to dynamic, generative systems. Historically, the integration of digital assistants began with reactive tools designed primarily for rapid information access and the execution of rudimentary, pre-defined commands (Kuhail et al., 2023). These early iterations operated within rigid, rule-based frameworks, which inherently constrained their capacity to address complex or divergent pedagogical queries.

However, the emergence of Generative Artificial Intelligence (GenAI) has fundamentally disrupted this paradigm. Modern AI chatbots have moved beyond simple command execution to the synthesis of sophisticated outputs, including text, code, and multimodal content. Unlike their predecessors, contemporary systems leverage deep learning architectures to facilitate fluid, context-aware dialogues, a capability that has significantly enhanced their efficacy within academic environments (Baidoo-Anu & Owusu Ansah, 2023). In the modern classroom, these assistants provide multifaceted support: they alleviate administrative burdens for educators through automated grading and content generation, while fostering student engagement via interactive and gamified learning experiences. Furthermore, these systems enable personalized tutoring, offering adaptive support for special education and granular feedback for language acquisition.

The core computational engines driving these advancements are Large Language Models (LLMs), characterized as sophisticated deep learning architectures trained on expansive datasets to generate human-like language (Touvron et al., 2023). While LLMs have revolutionized natural language processing, their deployment in education is often hindered by challenges such as the generation of factually inaccurate content and the limitations of static training data.

To address these constraints, recent scholarship has explored the utility of Small Language Models (SLMs). (Touvron et al., 2023). Models such as Gemma offer a compact alternative to high-parameter systems like Gemini or DeepSeek.. Having significantly fewer parameters, SLMs require substantially less computational power while maintaining competitive performance levels. Consequently, they represent an affordable and scalable solution for specialized educational tasks, particularly in environments where hardware resources are constrained.

A persistent obstacle in the pedagogical application of AI is the phenomenon of "hallucination," wherein a model generates factually incorrect information. To mitigate this risk, the Retrieval-Augmented Generation (RAG) paradigm has emerged as a robust architectural solution (Lewis et al., 2020). RAG functions by enabling a model to query a specific external knowledge base—such as verified curriculum textbooks—to retrieve factual, contemporary data prior to response generation. This approach bridges the gap between a model's generative capacity and the necessity for empirical accuracy.

According to Lewis et al. (2020), RAG effectively synthesizes "parametric memory," derived from pre-training, with "non-parametric memory," consisting of external document repositories. The technical workflow begins with indexing, where external data is segmented and converted into vector representations through embeddings. As Johnson et al. (2019) observe, the utilization of vector databases facilitates efficient similarity searches in high-dimensional spaces, allowing for the rapid retrieval of relevant context based on metrics such as cosine similarity. By utilizing retrieved data as a grounding context, the probability of hallucinations is drastically reduced, as the model's output is constrained by the real-time information provided. Shuster et al. (2021) demonstrate that incorporating such external knowledge not only improves response fidelity but also allows the system to cite its sources, thereby enhancing transparency and institutional trust. Consequently, RAG serves as a dynamic architecture that ensures AI systems remain current without the prohibitive costs of continuous fine-tuning.

To transition these theoretical frameworks into functional educational tools, deployment frameworks such as Streamlit[2] are frequently employed to transform Python-based RAG backends into accessible web applications. Such platforms facilitate complex user-interaction models, including "dual-persona" mechanisms. This allows a single underlying model to serve distinct roles—such as "Student" or "Teacher"—thereby providing role-specific functionalities within a unified interface (Zamfirescu-Pereira et al., 2023).

Building upon these advancements, this study introduces **S.T.A.R.T. BOT (Student Tutor Adaptive Role Tool Bot)**, a generative AI application tailored for the Greek educational sector. In our work, we use the aforementioned technologies to create a digital assistant leveraging a Retrieval-Augmented Generation (RAG) framework to ground responses in official textbooks. This architecture employs semantic segmentation to preserve curriculum integrity through metadata-enriched units. A core innovation is the Pedagogical Persona Switching mechanism, which utilizes persona-based prompting to allow a single model to pivot between an educator's assistant and a student's learning guide. Our pilot study in Home Economics evaluates if Small Language Models (SLMs) can offer a safe, cost-effective alternative to LLMs.

The remainder of this paper is organized as follows: The Methodology section delineates the framework and technical implementation of S.T.A.R.T. Bot. In the Experiments section, we detail the empirical evaluation of the chatbot's functionality and provide a comparative analysis between Large Language Models (LLMs) and a Small Language Model (SLM). The findings of this comparison are presented in the Results section, while the paper concludes with a summary of the research and its pedagogical implications.

## 3. Methodology

The methodology for implementing the system is divided into three main phases: the pre-processing of the data, the creation of the search index and the design of the interaction mechanisms.

### A. Data Pre-processing and Semantic Chunking

The process begins with the use of official school textbooks (Student's books) as primary data. To optimize recovery, the Chunking technique is applied:

1. The text is divided into conceptual units per chapter.

---

[2] https://streamlit.io/

2. Each section is limited to a size of up to 350 words, ensuring that the information remains focused and manageable by the model.
3. Each chunk is enriched with metadata, which includes the module, subsection and type of document, in order to maintain the structure of the curriculum.

**B. Semantic Indexing and Vector Storage**

To convert text into embeddings that allow semantic searching, the `lighteternal/stsb-xlm-r-greek-transfer` encoder from the `SentenceTransformers` library is used, which is specially trained for the Greek language.

- The generated vectors are stored in a FAISS index (specifically in the `faiss formula.IndexFlatL2`) for fast search based on Euclidean distance.
- This process allows the system to find the most relevant quotes from books based on the meaning of the user's question and not just the keywords.

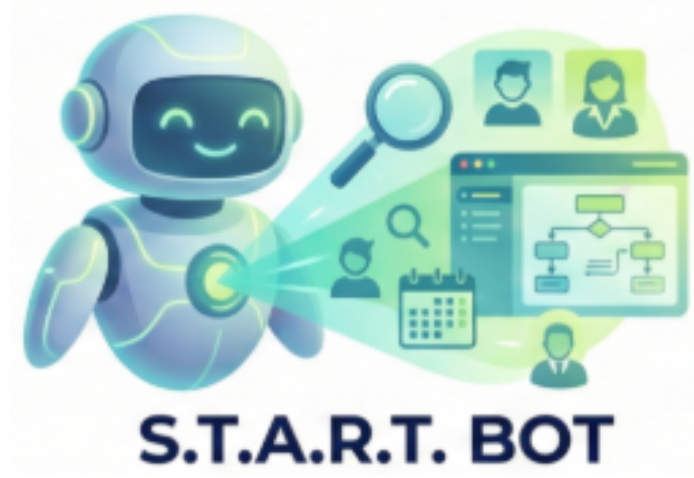

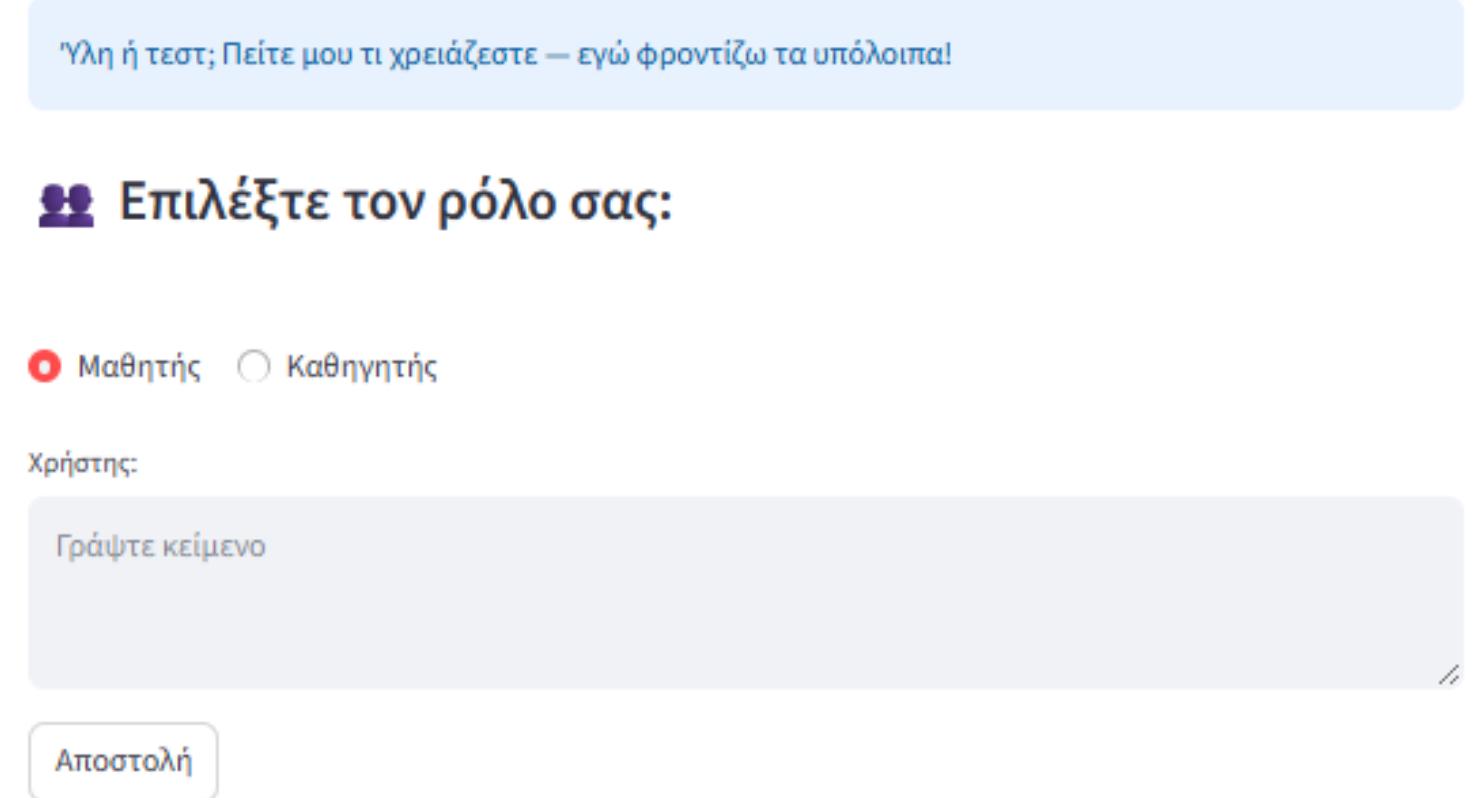

Figure 1: The S.T.A.R.T. BOT application using the Streamlit library

**C. Dual-Persona Prompting and Deployment**

The application is implemented with the **Streamlit** framework, offering an interactive web interface where the user chooses their role: **Student** or **Teacher**.

- **Persona Switching:** The architecture uses specialized Prompts that combine role, behavioral instructions, retrieved data (RAG), and response rules.
- **Student Mode:** Focuses on providing explanations and feedback based on the syllabus.

- **Tutor Mode:** Allows for the creation of educational materials, such as assessment tests and experiments.

The system supports the use of both large models (LLMs) such as `Gemini-1.5-Flash-Lite` and **DeepSeek,** as well as smaller models (SLMs) such as **Gemma 3:4b**, to evaluate their efficiency in an educational environment.

## 4. Experiments

The experimental phase of this study was designed to evaluate the pedagogical effectiveness and technical reliability of the S.T.A.R.T. BOT across its two distinct operational modes. The primary subject for the case study was Home Economics for Greek lower-secondary education, chosen for its cross-disciplinary nature and focus on practical life skills

### Language Models and Setup

To assess the impact of model scale on educational outcomes, the system utilized three different language models:

- Large Language Models (LLMs): Gemini (`gemini-1.5-flash-lite`) and DeepSeek (`deepseek-chat`).
- Small Language Model (SLM): Gemma (`gemma3:4b`), representing a resource-efficient alternative.

All models were integrated into a Retrieval-Augmented Generation (RAG) framework using a specialized Greek semantic encoder (lighteternal/stsb-xlm-r-greek-transfer) and a FAISS index for high-performance vector retrieval.

### Prompt Engineering and Persona Logic

A unified prompt structure was applied to maintain consistency across all experiments. The prompts followed a modular architecture:

**Prompt Structure:**

**“Role” + “Role Prompt” + “RAG Context” + “Rules” + “User Question”.**

This structure ensured that each model strictly adhered to its assigned pedagogical persona while grounding its responses in the retrieved educational segments from official textbooks.

### Student Mode Experiments

| Question | Gemma Answer |
|---|---|
| Να περιγράψετε την αγοραστική δραστηριότητα μιας οικογένειας. Να χρησιμοποιήσετε ένα παράδειγμα | Σύμφωνα με τα αποσπάσματα, η αγοραστική δραστηριότητα μιας οικογένειας περιλαμβάνει διάφορα στάδια. Πριν αγοράσει ένα προϊόν, η οικογένεια διαπιστώνει την ανάγκη που πρέπει να ικανοποιήσει, αναζητά πληροφορίες για τα διάφορα προϊόντα, διερευνά τις διάφορες αγορές από τις οποίες μπορεί να προμηθευτεί το προϊόν, συγκρίνει τις τιμές και γενικότερα τις διάφορες προσφορές, αξιολογεί όλες τις προτάσεις, και λαμβάνει την τελική απόφαση. Για παράδειγμα, αν μια |

| **Gold Answer** | οικογένεια επιθυμεί να αγοράσει ένα αυτοκίνητο, πρέπει αρχικά να συλλέξει πληροφορίες για τους διάφορους τύπους και μάρκες αυτοκινήτων, να συγκρίνει τις τιμές των αυτοκινητοβιομηχανιών και, αφού τις αξιολογήσει, να αποφασίσει ποιο αυτοκίνητο θα αγοράσει τελικά. Επίσης, η οικογένεια έχει βασικές πηγές εισοδήματος (μισθό, ελευθέριο επάγγελμα) και καταρτίζει έναν οικογενειακό προϋπολογισμό για να διαχειριστεί τα έσοδά της και να αποφασίσει πώς θα ξοδέψει τα χρήματά της.<br>📚 Πηγές από την ύλη που χρησιμοποιήθηκαν:<br>🔸 21-0035-02_Oikiaki-Oikonomia_A-Gymnasiou_Vivlio-Mathiti.pdf \| 2.2.3 — Οικογενειακός προϋπολογισμός<br>🔸 21-0035-02_Oikiaki-Oikonomia_A-Gymnasiou_Vivlio-Mathiti.pdf \| 2.2.2 — Ο οικονομικός ρόλος της οικογένειας<br>🔸 21-0095-03_Oikiaki-Oikonomia_A-B-Gymnasiou_Vivlio-Mathiti.pdf \| 4.4.6 — είναι «άπιαστα». |
|---|---|
| Η αγοραστική δραστηριότητα μιας οικογένειας έχει ως εξής: διαπίστωση της ανάγκης, αναζήτηση των κατάλληλων πληροφοριών, έρευνα των διάφορων αγορών και τέλος σύγκριση των τιμών και γενικότερα των διάφορων προσφορών. Ως παράδειγμα οι μαθητές θα μπορούσαν να αναφέρουν την αγορά ενός αυτοκινήτου. | |

Figure 2: An example of Student mode S.T.A.R.T. BOT response – model `gemma3:4b`

In the Student Mode, the primary objective was to measure the accuracy of curriculum-grounded explanations.

- Execution: A set of 10 curriculum-based questions was presented to each of the three models.
- Evaluation: The generated answers were compared against "gold answers" extracted from the official teacher's guide.
- Metrics: Semantic similarity scores were calculated between the models' outputs and the reference answers to quantify factual and conceptual alignment.

**Tutor Mode Experiments**

In the Tutor Mode, the focus shifted to the assistant's ability to act as an instructional partner by generating complex pedagogical artifacts.

- Execution: Each model was requested to generate fully structured assessment tests.
- Scope: Two different chapters from the Home Economics curriculum were selected for test generation (e.g., "Family Budget" and "Modern Technology and Food").
- Evaluation Criteria: The generated tests were evaluated based on three key metrics: coverage rate of the instructional units, quality of question formulation for junior high school students, and overall relevance to the provided educational material

| Question | Gemini Answer |
|---|---|

Θέλω να δημιουργήσεις ένα σύντομο αλλά πλήρως δομημένο σετ ερωτήσεων αξιολόγησης που καλύπτει την ύλη του Κεφαλαίου 5.12 Σύγχρονη τεχνολογία και τρόφιμα ως εξής:
Α. Ανοιχτού τύπου: 2–3 ερωτήσεις
Β. Πειράματα: Περιγραφή 2–3 πειραμάτων για την συντήρηση των τροφίμων

Η Τεχνητή Νοημοσύνη συναντά την τάξη!

Ύλη ή τεστ; Πείτε μου τι χρειάζεστε — εγώ φροντίζω τα υπόλοιπα!

Επιλέξτε τον ρόλο σας:

Μαθητής Καθηγητής

Χρήστης:

Θέλω να δημιουργήσεις ένα σύντομο αλλά πλήρως δομημένο σετ ερωτήσεων αξιολόγησης που καλύπτει την ύλη του Κεφαλαίου 5.12 Σύγχρονη τεχνολογία και τρόφιμα ως εξής:
Α. Ανοιχτού τύπου: 2–3 ερωτήσεις
Β. Πειράματα: Περιγραφή 2–3 πειραμάτων για την συντήρηση των τροφίμων

Αποστολή

Βοηθός:

Α. Ερωτήσεις Ανοιχτού Τύπου:

1. Περιέγραψε γιατί είναι σημαντικό να καταναλώνουμε ποικιλία τροφίμων καθημερινά.
2. Πώς η γνώση της σύστασης των τροφίμων, όπως αυτή αναγράφεται στις ετικέτες, βοηθά τον καταναλωτή να κάνει υγιεινότερες επιλογές;
3. Εξήγησε γιατί οι σύγχρονες μέθοδοι συντήρησης των τροφίμων ενισχύουν τις δυνατότητες επιλογών του καταναλωτή.

Β. Πειράματα για τη Συντήρηση των Τροφίμων:

1. **Πείραμα: Επίδραση της Ψύξης στη Συντήρηση του Γάλακτος**
   - **Υλικά:** Δύο γυάλινα ποτήρια, γάλα, δύο καπάκια.
   - **Διαδικασία:**
     - Γεμίστε το ένα ποτήρι με γάλα και καλύψτε το με καπάκι.
     - Γεμίστε το δεύτερο ποτήρι με γάλα, χωρίς να το καλύψετε.
     - Αφήστε το πρώτο ποτήρι στο ψυγείο και το δεύτερο σε θερμοκρασία δωματίου.
   - **Παρατήρηση:** Μετά από 24 ώρες, παρατηρήστε και καταγράψτε τυχόν αλλαγές στην εμφάνιση και τη μυρωδιά του γάλακτος και στα δύο ποτήρια. Συζητήστε γιατί παρατηρήθηκαν αυτές οι διαφορές.

Figure 3: An example of Tutor mode S.T.A.R.T. BOT response – model `gemini-1.5-flash-lite`

## 5. Evaluation and Results

### Student Mode

The evaluation in **Student Mode** centered on measuring the accuracy and semantic consistency of the generated responses relative to the official curriculum. The "**gold answers**" extracted from the official teacher's guide for the Home Economics course served as the ground truth for this assessment.

### Semantic Similarity Analysis

Quantitative evaluation was conducted using Semantic Similarity metrics. The results were derived from a comparative study of 10 curriculum-based questions submitted to the three models: Gemini, DeepSeek, and the SLM Gemma.

### Key Findings

- The small language model **Gemma** demonstrated significant competitiveness, achieving a mean score of **0.753**. This performance is remarkably close to the larger models, Gemini (0.773) and DeepSeek (0.764).
- In several instances, such as Q1 and Q3, Gemma outperformed the LLMs, reaching similarity scores as high as **0.914**. This highlights the efficacy of SLMs when supported by a well-structured RAG architecture.
- Qualitative analysis confirmed that the system maintained high fidelity to the official content. For example, when asked to describe a family's purchasing activity, Gemma produced a comprehensive response covering all curriculum-specified stages—needs identification, information search, and price comparison—while providing precise citations to the source textbooks

| Question | Gemini | DeepSeek | Gemma |
|---|---|---|---|
| 1 | 0,741 | 0,776 | **0,856** |
| 2 | **0,939** | 0,928 | 0,875 |
| 3 | 0,811 | 0,815 | **0,914** |
| 4 | 0,586 | 0,574 | **0,640** |
| 5 | 0,650 | **0,738** | 0,588 |
| 6 | **0,883** | 0,756 | 0,698 |
| 7 | 0,886 | 0,872 | **0,894** |
| 8 | **0,900** | 0,865 | 0,822 |
| 9 | 0,755 | **0,779** | 0,679 |
| 10 | **0,575** | 0,537 | 0,561 |
| **Mean** | **0,773** | **0,764** | **0,753** |

Table 1. Student Mode - Semantic Similarity Results

**Tutor Mode (LLM-as-a-Judge).**

**Evaluation Methodology**
To ensure an objective and scalable assessment test generated in Tutor Mode, the study employed the **LLM-as-a-Judge** framework. The state-of-the-art model **GPT-5.1 pro** was utilized as an expert to evaluate the assessment tests produced by the three candidate models (Gemini, DeepSeek, and Gemma).

The evaluation process was structured around a specialized "**Evaluation Prompt**" that instructed the judge model to analyze the generated content based on three key dimensions:

1. **Coverage Rate:** Examination of the extent to which the generated test covers the core concepts of the specific instructional unit.
2. **Relevance to Material:** Evaluation of how closely the questions align with the provided RAG context from the Home Economics textbooks. Verification of the artifact's format, including the logical flow of questions and the inclusion of diverse assessment types (e.g., experiments, open-ended questions).
3. **Quality of question formulation** in relation to comprehension by a high school student.

Our study minimized human bias and provided a standardized comparison across different model architectures. The judge model provided both qualitative feedback and quantitative scoring, allowing for a rigorous analysis of how effectively each model can serve as an instructional partner for educators.

**Performance Comparison and Model Reliability**
A significant finding during the experimental phase was the complete failure of the DeepSeek model to generate any output for the requested assessment tests in either of the two selected chapters. This lack of responsiveness indicates a potential limitation in the model's reliability or its ability to handle complex, persona-driven instructions within the Greek educational context. Consequently, the comparative analysis focused on the performance of Gemini and the SLM Gemma.

| TEST 1 | Coverage rate | Quality of the formulation of the questions | Relevance to the educational material |
|---|---|---|---|
| Gemini | 65 – 75 % | Οι ερωτήσεις είναι γενικά **καλογραμμένες, σαφείς και σε σωστό γλωσσικό επίπεδο** για Γυμνάσιο.<br>Το μόνο πραγματικό παιδαγωγικό «γκρίζο» σημείο είναι η **Α3**, λόγω αδυναμίας σύνδεσης με το συγκεκριμένο απόσπασμα και υψηλότερου επιπέδου αφαίρεσης. | Η πλειονότητα των ερωτήσεων είναι σαφώς σχετική και «πατάει» πάνω στο κείμενο.<br>Μικρή «διαρροή» σε άλλη υποενότητα (Γ4) και μία ερώτηση πιο γενική (Α3). |
| Gemma | 60 – 70 % | Η γλωσσική ποιότητα είναι **καλή** και φιλική προς τον μαθητή.<br>Γενικά καθαρή και κατάλληλη για μαθητές Γυμνασίου, με 2–3 σημεία που χρειάζονται καλύτερη σύνδεση με τη φρασεολογία του βιβλίου. | Οι περισσότερες ερωτήσεις είναι καλά ευθυγραμμισμένες, αλλά ορισμένες (κυρίως Α2 & Α4) πατούν πιο πολύ σε «λογική/εμπειρική» γνώση παρά σε ρητό κείμενο. |

Table 2. Assessment Test 1: Family Budget

**Assessment Test 1: Family Budget**
In the first test, which covered the "Family Budget" unit, both active models demonstrated strong pedagogical alignment. Gemini achieved a coverage rate of 65% – 75%, producing clear and linguistically appropriate questions for the target age group. It showed superior overall coverage in sub-units related to budgeting, income, and planning. Gemma (SLM) followed closely with a 60% – 70% coverage rate, maintaining high linguistic quality and a student-friendly tone. Notably, Gemma excelled in generating items related to payment methods, though it occasionally relied on general logic rather than explicit textbook phrasing.

While the BookTest1 (derived directly from the textbook) remained the "gold standard" with 100% compatibility, Gemini was identified as the top-performing AI model for this specific unit due to its structured academic phrasing and minimal deviations from the curriculum.

| TEST 2 | Coverage rate | Quality of the formulation of the questions | Relevance to the educational material |
|---|---|---|---|
| Gemini | 50 – 60 % | Γλωσσικά καθαρό και κατανοητό. Τα πειράματα είναι καλά ως μορφή, αλλά το 2ο δεν αναδεικνύει σωστά την αποστείρωση. Δύο ανοικτές ερωτήσεις (ποικιλία, ετικέτες) | Η πλειονότητα των ερωτήσεων είναι σαφώς σχετική και «πατάει» πάνω στο κείμενο. Μικρή «διαρροή» σε άλλη υποενότητα (Γ4) και μία ερώτηση πιο γενική (Α3). |

|  |  | δεν «κουμπώνουν» με το συγκεκριμένο κείμενο. |  |
|---|---|---|---|
| Gemma | 60 – 70 % | Πολύ ικανοποιητική – γλώσσα απλή, σαφής, με ξεκάθαρο τι ζητείται και με καλή δομή στα πειράματα. | Καλή μόνο για μέρος του υλικού (σημασία συντήρησης, σύγχρονες μέθοδοι, δύο βασικές μέθοδοι ψύξης/ξήρανσης). Αδύναμη έως χαμηλή για αρκετά βασικά στοιχεία (ετικέτες, ειδικές μέθοδοι, μικροοργανισμοί/ ένζυμα, συντηρητικά, ζύμωση). |

Table 3. Assessment Test 2: Modern Technology and Food

**Assessment Test 2: Modern Technology and Food**
The second test focused on more technical content regarding food preservation and technology. Gemma (SLM) emerged as the superior AI performer in this category, achieving a 60% – 70% coverage rate and receiving high praise for its simple, clear language and well-structured experimental descriptions. It provided an excellent pedagogical transcription of theory into practical experiments. Gemini showed a lower coverage rate of 50% – 60%. While its language was clear, the evaluator noted that its questions tended toward general nutritional advice rather than the specific technical details of the instructional unit, such as food labeling or specific preservation methods.

**Summary of Findings:** The results from the Tutor Mode highlight a nuanced trade-off between model architectures. While Gemini provided a more "academic" and strictly aligned structure for economic concepts, the SLM Gemma demonstrated superior pedagogical creativity, particularly in designing classroom experiments and maintaining a friendly tone for young learners. Crucially, the experiments confirm that Small Language Models can effectively serve as instructional assistants, often rivaling or exceeding the performance of larger models in specific creative teaching tasks.

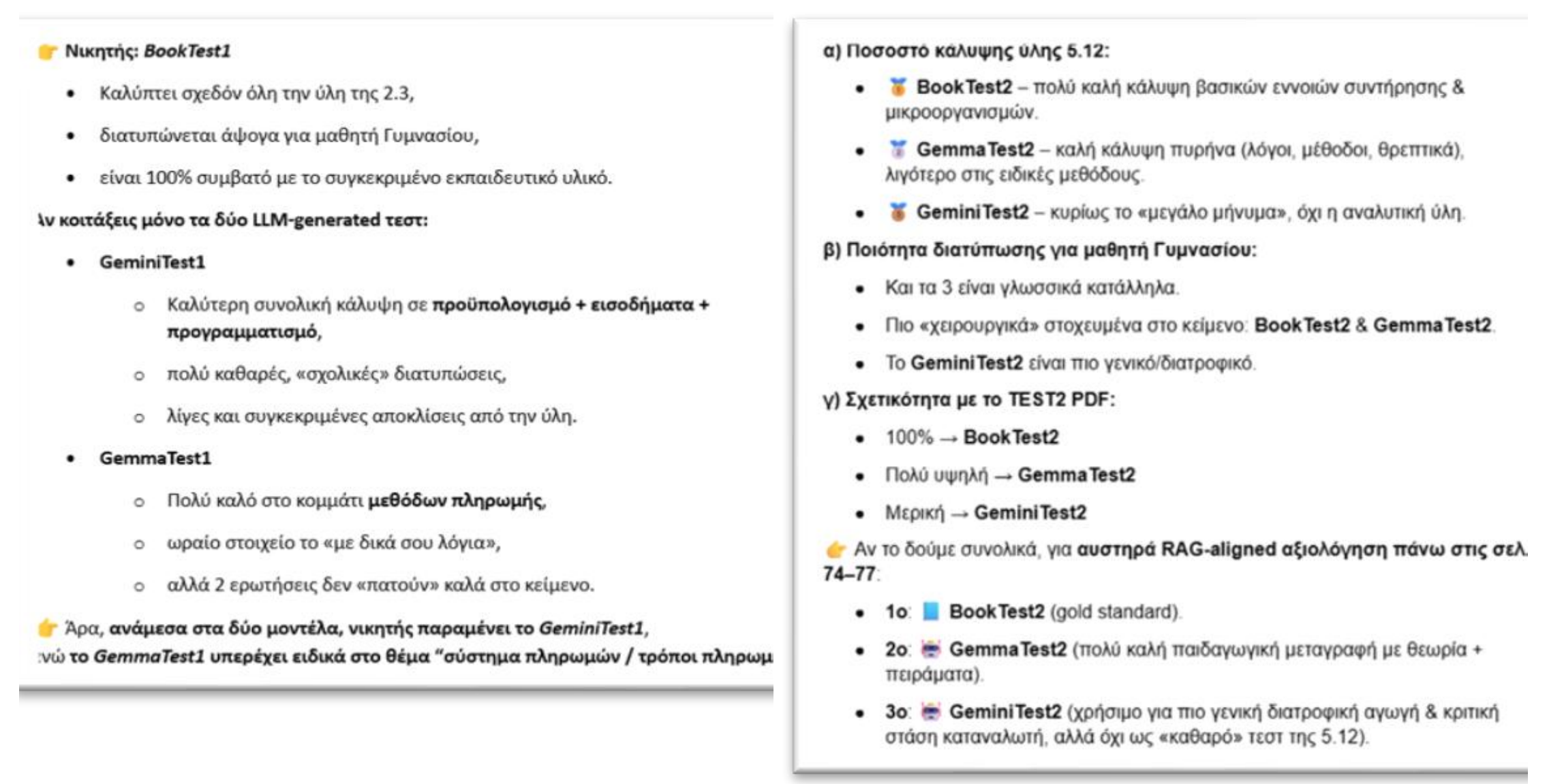

Figure 4: GhatGPT as a Judge - Summary of Findings

## 6. Conclusion

The development and evaluation of the S.T.A.R.T. BOT demonstrate that the integration of Generative AI within a structured Retrieval-Augmented Generation (RAG) framework can significantly enhance the educational process in the Greek secondary sector. The study successfully implemented a Pedagogical Persona Switching mechanism, proving that a single AI model can effectively transition between the roles of a learning guide for students and an instructional partner for teachers. Regarding the core research questions, the experimental results provide compelling evidence that Small Language Models, such as Gemma, can decisively compete with significantly larger models like Gemini and DeepSeek in specialized educational tasks. In both Student and Tutor modes, the SLM demonstrated high semantic alignment with the official curriculum, often matching or even exceeding the performance of LLMs in pedagogical creativity and linguistic clarity.

A primary conclusion of this research is the distinct suitability of SLMs for deployment within closed educational environments. Unlike LLMs, which typically require cloud-based processing and API calls to external servers, SLMs can be hosted on-premise or on local school infrastructure. This architectural advantage ensures that sensitive educational data and student interactions remain local, eliminating the risk of information leakage to the public internet and aligning with strict data privacy and GDPR requirements.

Furthermore, the study highlights the efficiency of SLMs in terms of reduced hardware requirements. While LLMs demand immense computational power and high-end GPU clusters, SLMs are designed to operate effectively with significantly fewer resources. This low computational footprint makes them a sustainable and cost-effective solution for large-scale integration into national education systems, where hardware budgets may be constrained.

In conclusion, the S.T.A.R.T. BOT framework establishes a reliable, secure, and pedagogically grounded foundation for the future of AI in education. By leveraging the power of SLMs and RAG, we can provide teachers and students with advanced digital assistants that are not only intelligent and curriculum-aligned but also private, localized, and technologically accessible.

**Contact email:** cmizeli@ihu.edu.gr